\documentclass[12pt]{iopart}
\usepackage{natbib}

\usepackage{listings}
\usepackage{float}
\newfloat{listing}{thHp}{lop}
\newcommand{\pythoncode}{\lstinputlisting}

\usepackage{hyperref}

\usepackage{graphicx}
\graphicspath{{./images/}}

\usepackage{perpage} %the perpage package
\MakePerPage{footnote} %the perpage package command

\usepackage{aas_macros}

\begin{document}

\title{SunPy - Python for Solar Physics}

\author{The SunPy Community,}
\address{\url{http://sunpy.org}}
\ead{sunpy@googlegroups.com}

\author{
% People who have directly contributed to the text, still in no order.
Stuart J Mumford$^1$, 
Steven Christe$^2$, 
David P\'erez-Su\'arez$^3$, 
Jack Ireland$^{2,4}$, 
Albert Y Shih$^2$, 
Andrew R Inglis$^{2,5}$, 
Simon Liedtke$^{6}$, 
Russell J Hewett$^7$, %88
% Contributors to SunPy Sorted thus:
% No. Commits to main repo + commits in open PR's
% Date of most recent contribution
Florian Mayer$^{8}$, %721
Keith Hughitt$^9$, %446
Nabil Freij$^1$, %52
Tomas Meszaros$^{10}$, %17
Samuel M Bennett$^1$, %14
Michael Malocha$^{11}$, %10
John Evans$^{12}$, %10 %No affil
Ankit Agrawal$^{13}$, %8
Andrew J Leonard$^{14}$, %3
Thomas P Robitaille$^{15}$, %2
Benjamin Mampaey$^{16}$, %2
Jose Iv\'an Campos-Rozo$^{17}$, %1
and Michael S Kirk$^{2}$
}

\address{
$^1$Solar Physics \& Space Plasma Research Centre (SP$^{2}$RC), 
School of Mathematics and Statistics, The University of Sheffield, Hicks 
Building, Hounsfield Road, Sheffield, S3 7RH, UK

$^2$NASA Goddard Space Flight Center, Greenbelt, MD, USA

$^3$South African National Space Agency - Space Science, Hospital 
Street, 7200 Hermanus, Western Cape, South Africa

$^4$ADNET Systems Inc., Mail Code 671.1, NASA Goddard Space Flight Center, 
Greenbelt, MD, USA

$^5$The Catholic University of America, Washington, DC, USA

$^{6}$University of Bremen, Bibliothekstra\ss e 1, 28359 Bremen, Germany

$^7$Department of Mathematics, Massachusetts Institute of Technology, 77 
Massachusetts Ave, E17-317, Cambridge, MA, USA

$^{8}$Vienna University of Technology, Karlsplatz 13, 1040 Vienna, Austria

$^9$Department of Cell Biology and Molecular Genetics, University of Maryland, College Park, MD, USA

$^{10}$Masaryk University, Faculty of Informatics, Botanicka 68a Brno, Czech Republic

$^{11}$Humboldt State University, 1 Harpst St, Arcata, CA, USA

$^{12}$Boston Python User Group, Boston, MA, USA

$^{13}$Indian Institute of Technology, Bombay, India

$^{14}$Department of Mathematics and Physics, Aberystwyth University, 
Physical Sciences Building, Aberystwyth, SY23 3BZ, UK

$^{15}$Max-Planck-Institut f\"{u}r Astronomie, K\"onigstuhl 17, Heidelberg
69117, Germany

$^{16}$Royal Observatory of Belgium, Brussels, Belgium

$^{17}$Observatorio Astron\'omico Nacional, Universidad Nacional de Colombia, 
Bogot\'a, D.C., Colombia

}

\begin{abstract}
This paper presents SunPy (version 0.5), a community-developed Python package 
for solar physics.
Python, a free, cross-platform, general-purpose, high-level programming 
language, has seen widespread adoption among the scientific community, resulting 
in the availability of a large number of software packages, from numerical 
computation (\texttt{NumPy}, \texttt{SciPy}) and machine learning (\texttt{scikit-learn}) to visualisation 
and plotting (\texttt{matplotlib}).
SunPy is a data-analysis environment specialising in providing the software 
necessary to analyse solar and heliospheric data in Python. 
SunPy is open-source software (BSD licence) and has an open and transparent 
development workflow that anyone can contribute to.
SunPy provides access to solar data through integration with the Virtual 
Solar Observatory (VSO), the Heliophysics Event Knowledgebase (HEK), and the 
HELiophysics Integrated Observatory (HELIO) webservices. It currently supports image data from major solar missions (e.g., 
\textit{SDO}, \textit{SOHO}, \textit{STEREO}, and \textit{IRIS}), time-series data from 
missions such as \textit{GOES}, \textit{SDO}/EVE, and \textit{PROBA2}/LYRA, and radio 
spectra from e-Callisto and \textit{STEREO}/SWAVES. We describe SunPy's functionality,
provide examples of solar data analysis in SunPy, and show how Python-based solar 
data-analysis can leverage the many existing tools already 
available in Python. We discuss the future goals of the project and encourage 
interested users to become involved in the planning and development of SunPy.
\end{abstract}

\maketitle

\section{Introduction}\label{sec:Intro}

Science is driven by the analysis of data of ever-growing variety and 
complexity. Advances in sensor technology, combined with the availability of 
inexpensive 
storage, have led to rapid increases in the amount of data available to scientists in almost
every discipline.  Solar physics is no exception to this trend. For example,
NASA's \textit{Solar Dynamics Observatory} (\textit{SDO}) spacecraft, launched
in February 2010, produces over 1 TB of data per day \citep{pesnell2012}. Managing and
analysing these data requires increasingly sophisticated software
tools. These tools should be robust, easy to use and modify, have a transparent
development history, and conform to modern software-engineering
standards. Software with these qualities provide a strong foundation that can support the
needs of the community as data volumes grow and science questions evolve.

%The SunPy project aims to provide a free, open-source, and openly developed
%software package for the analysis and visualisation of solar data.
The SunPy project aims to provide a software package with these qualities for 
the analysis and visualisation of solar data. SunPy makes
use of Python and scientific Python packages. Python is a free, general-purpose, 
powerful, and easy-to-learn high-level programming language. Additionally, Python is 
widely used outside of scientific fields in areas such as `big data' analytics, web 
development, and educational environments. For example, \texttt{pandas} \citep{mckinney2010, mckinney2012} was 
originally developed for quantitative analysis of financial data and has since 
grown into a generalised time-series data-analysis package. Python continues to 
see increased use in the astronomy community \citep{greenfield2011}, which has 
similar goals and requirements as the solar physics community. Finally, Python 
integrates well with many technologies such as web servers \citep{dolgert2008} and databases. 

The development of a package such as SunPy is made possible by the rich ecosystem of 
scientific packages available in Python. Core packages such as \texttt{NumPy}, 
\texttt{SciPy} \citep{jones2001}, and \texttt{matplotlib} \citep{hunter2007} provide 
the basic functionality expected of a scientific programming language,
such as array manipulation, core numerical algorithms, and visualisation, respectively.
Building upon these foundations, packages such as \texttt{astropy}
\citep[astronomy;][]{theastropycollaboration2013}, \texttt{pandas} \citep[time-series;][]{mckinney2012}, and \texttt{scikit-image} 
\citep[image processing;][]{vanderwalt2014} provide more domain-specific functionality.

A typical workflow begins with a solar physicist manually identifying
a small number of events of interest on the Sun.  This is typically
done in order to investigate in detail the physics of these events
(for example, the large solar flare of 23 July 2002 has Astrophysical
Journal Letters volume 595, dedicated to its analysis).
In this workflow, an event is investigated in depth which requires 
data from many different instruments.
These data are typically provided in many different formats - for
example, FITS \citep[Flexible Image Transport System,][]{refId0}, CSV, or
binary files - and contain many different types of data (such as
images, lightcurves and spectra).  In addition, the repositories these data reside
in can have different access methods.  This workflow is characterized
by the large number of heterogeneous datasets used in the
investigation of a small number of solar events.

Another typical workflow begins with the solar physicist identifying a
large sample of data or events.  The goal here is obtain information
about the population in general.  An example might be to calculate the
fractal dimension of a large number of active region magnetic fields
\citep{2005ApJ...631..628M}, or to calculate the observed temperatures
in a population of solar flares \citep{2012ApJS..202...11R}.  This
workflow is typically characterized by lower data heterogeneity, but
with a larger number of files.

The volume and variety of solar data used in these workflows drives
the need for an environment in which obtaining and performing common
solar physics operations on these data is as simple and intuitive as
possible.  SunPy is designed to be a clean, simple-to-use, and
well-structured open-source package that provides the \textit{core}
tools for solar data analysis, motivated by the need for a free and
modern alternative to the existing SolarSoft (SSW) library
\citep{freeland1998}. While SSW is open source and freely available,
it relies on IDL (Interactive Data Language), a proprietary
data-analysis environment.

The purpose of this paper is to provide an overview of SunPy's current
capabilities, an overview of the project's development model, community aspects of the
project, and future plans. The latest release of SunPy, version 0.5,
can be downloaded from \url{http://sunpy.org} or can be
installed using the Python package index (\url{http://pypi.python.org/pypi}).

\section{Core Data Types}\label{sec:DataTypes}

The core of SunPy is a set of data structures that are specifically
designed for the three primary varieties of solar physics data:
images, time series, and spectra. These core data types are supported
by the SunPy classes: \texttt{Map} (2D spatial data),
\texttt{LightCurve} (1D temporal series), and \texttt{Spectrum} and
\texttt{Spectrogram} (1D and 2D spectra).  The purpose of these
classes is to provide the same core data type to the SunPy user
regardless of the differences in source data.  For example, if two
different instruments use different time formats to describe the
observation time of their images, the corresponding SunPy \texttt{Map}
object for each of them expresses the observation time in the same
way.  This simplifies the workflow for the user when handling data
from multiple sources. 

These classes allow access to the data
and associated metadata and provide appropriate convenience functions to
enable analysis and visualisation. For each of these classes, the data is
stored in the \texttt{data} attribute, while the metadata is stored 
in the \texttt{meta} attribute\footnote{Note, that currently only \texttt{Map} and \texttt{LightCurve} have this feature 
fully implemented.}. 
It is possible to instantiate the
data types from various
different sources: e.g., files, URLs, and arrays.  
In order to provide instrument-specific specialisation, the core SunPy classes 
make use of subclassing; e.g., \texttt{Map} has an \texttt{AIAMap} 
sub-type for data from the \textit{SDO}/AIA (Atmospheric Imaging Assembly; \citealt{lemen2012}) instrument. 

All of the core SunPy data types 
include visualisation methods that are tailored to each data type. 
These visualisation methods all utilise the \texttt{matplotlib} 
package and are designed in such a way that they integrate well with 
the \texttt{pyplot} functional interface of \texttt{matplotlib}.

This design philosophy makes the behaviour of SunPy's visualisation 
routines intuitive to those who already understand the \texttt{matplotlib}
interface, as well as allowing the use of the standard 
\texttt{matplotlib} commands to manipulate the plot parameters (e.g., title, axes).
Data visualisation is provided by two functions: \texttt{peek()}, for quick 
plotting, and \texttt{plot()}, for plotting with more fine-grained control.

This section will give a brief overview of the \textit{current} functionality 
of each of the core SunPy data types.

\subsection{Map}\label{ssec:map}
The map data type stores 2D spatial data, such as images of the Sun and 
inner heliosphere. It provides: a wrapper around a \texttt{numpy} data array, 
the images associated spatial coordinates, and other metadata. The \texttt{Map} 
class provides methods for typical operations on 2D data, such as rotation and 
re-sampling, as well as visualisation.
The \texttt{Map} class also provides a convenient interface for loading data 
from a variety of sources, including from FITS files, the standard format for storing image data in solar physics and astrophysics community. 
An example of creating a \texttt{Map} object from a FITS file is shown in 
Listing~\ref{code:aia_1}.

The architecture of the map subpackage consists of a template map called
\texttt{GenericMap}, which is a subclass of \texttt{astropy.nddata.NDData}. 
\texttt{NDData} is a generic wrapper around a \texttt{numpy.ndarray} with a 
\texttt{meta} attribute to store metadata.
As \texttt{NDData} is currently still in development, \texttt{GenericMap} does 
not yet make full use of its capabilities, but this inheritance structure 
provides for future integration with \texttt{astropy}. In order to provide 
instrument- or detector-specific integration, \texttt{GenericMap} is designed
to be subclassed. Each subclass of \texttt{GenericMap} can register 
with the \texttt{Map} creation factory, which will then automatically return an instance
of the specific \texttt{GenericMap} subclass dependent upon the data provided. 
SunPy v0.5 has \texttt{GenericMap} specialisations for the following 
instruments: 

\begin{itemize}
\item \textit{Yohkoh} Solar X-ray Telescope (SXT, \citealp{1991SoPh..136....1O, 1991SoPh..136...37T}),
\item Solar and Heliospheric Observatory (\textit{SOHO}, \citealp{domingo1995}) Extreme Ultraviolet Telescope (EIT; \citealp{1995SoPh..162..291D})
\item \textit{SOHO} Large Angle Spectroscopic COronagraph (LASCO, \citealp{1995SoPh..162..357B})
\item \textit{RHESSI} - Reuven Ramaty High Energy Solar Spectroscopic Imager \citep{2002SoPh..210....3L},
\item Solar TErrestrial RElations Observatory (\textit{STEREO}, \citealp{2005AdSpR..36.1483K}) Extreme Ultraviolet Imager (EUVI, \citep{2004SPIE.5171..111W})
\item \textit{STEREO} CORonagraph 1/2 (COR 1/2, \citealp{2002AdSpR..29.2017H})
\item \textit{Hinode} XRT - X-Ray Telescope \citep{2007SoPh..243....3K, 2007SoPh..243...63G}.
\item PRojects for On Board Autonomy 2 (\textit{PROBA2}, \citealp{2013SoPh..286....5S}) Sun Watcher Active Pixel (SWAP; \citealp{2013SoPh..286...43S})
\item \textit{SDO} AIA and Helioseismic Magnetic Imager, (HMI, \citealp{2012SoPh..275..207S})
\item Interface Region Imaging Spectrograph (\textit{IRIS}, \citealp{2011SPD....42.1512L}) SJI (slit-jaw imager) frames.           
\end{itemize}
             
The \texttt{GenericMap} class stores all of the metadata retrieved from the header of
the image file in the \texttt{meta} attribute and provides convenience 
properties for commonly accessed metadata: e.g., \texttt{instrument}, 
\texttt{wavelength} or \texttt{coordinate\_system}.
These properties are dynamic mappings to the underlying metadata and all methods 
of the \textit{GenericMap} class modify the meta data where needed.
For example, if \verb|aiamap.meta[`instrume']| is modified then \verb|aiamap.instrument| 
will reflect this change.
Currently this is implemented by not preserving the keywords of the input data,
instead modifying meta data to a set of ``standard" keys supported by SunPy.
Listing \ref{code:aia_1} demonstrates the quick-look functionality of 
\texttt{Map}.

\begin{listing}[H]
\pythoncode{pycode_map1.txt}
\begin{center}
\includegraphics[width=0.8\columnwidth]{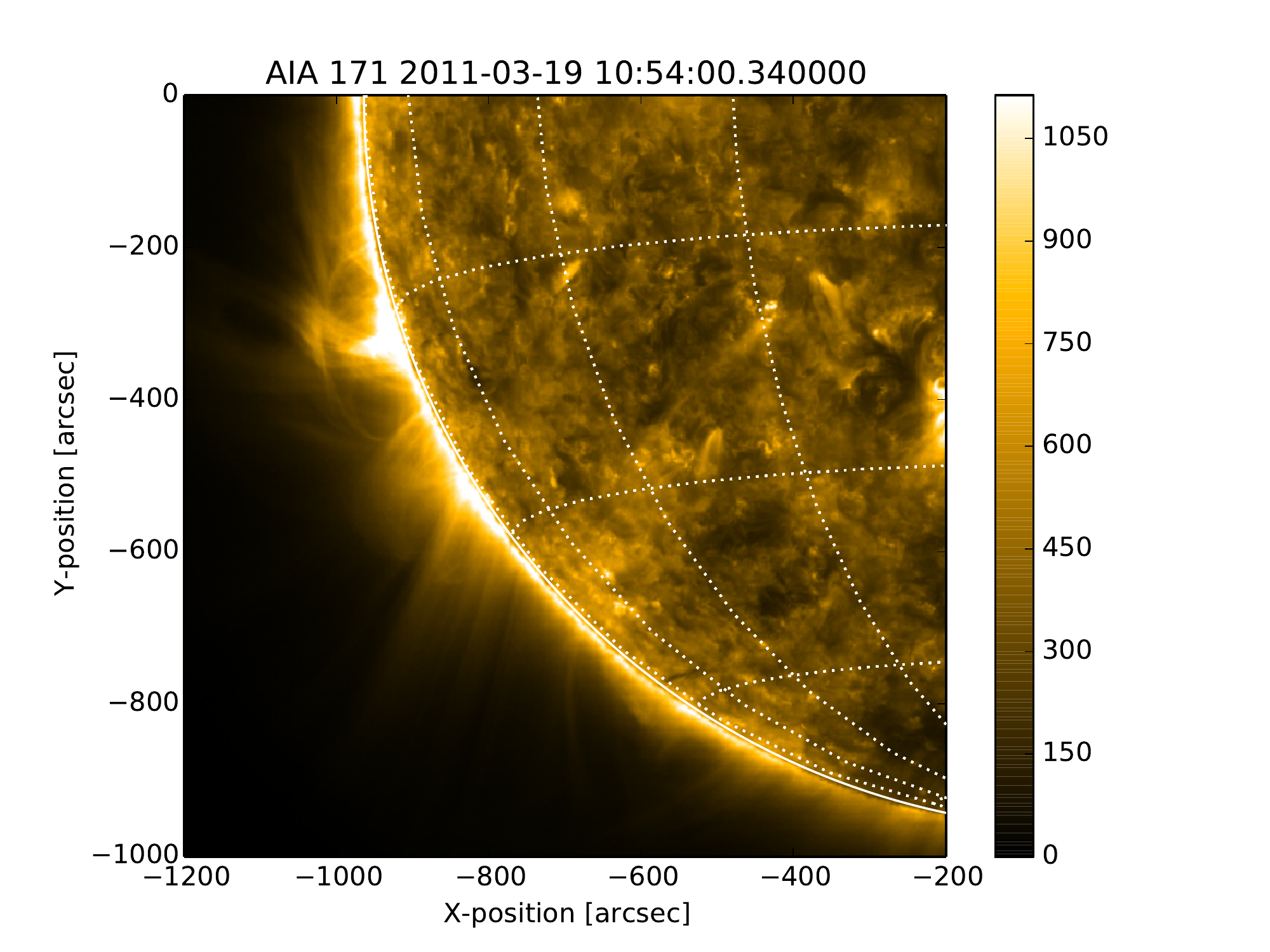}
\end{center}
\caption{Example of the \texttt{AIAMap} specialisation of 
\texttt{GenericMap}. First, a map is created from a sample \textit{SDO}/AIA FITS file. In this case, a demonstration file contained within the SunPy repository is used. A cutout
of the full map is then created by specifying the desired solar-$x$ and solar-$y$ ranges of the plot in data coordinates (in this case, arcseconds), and then a quick-view plot is created with lines of heliographic longitude and latitude over-plotted.}
\label{code:aia_1}
\end{listing}

In addition to the data-type classes, the \texttt{map} subpackage provides two 
collection classes, \texttt{CompositeMap} and \texttt{MapCube}, for 
spatially and temporally aligned data respectively.
\texttt{CompositeMap} provides methods for overlaying spatially aligned 
data, with support for visualisation of images and contour lines overlaid 
upon each other.
\texttt{MapCube} provides methods for animation of its series of \texttt{Map} 
objects. Listings~\ref{code:compmap_1} and \ref{code:mapcube_1} show how to 
interact with these classes.

\begin{listing}[H]
\pythoncode{pycode_map2.txt}
\begin{center}
\includegraphics[width=0.8\columnwidth]{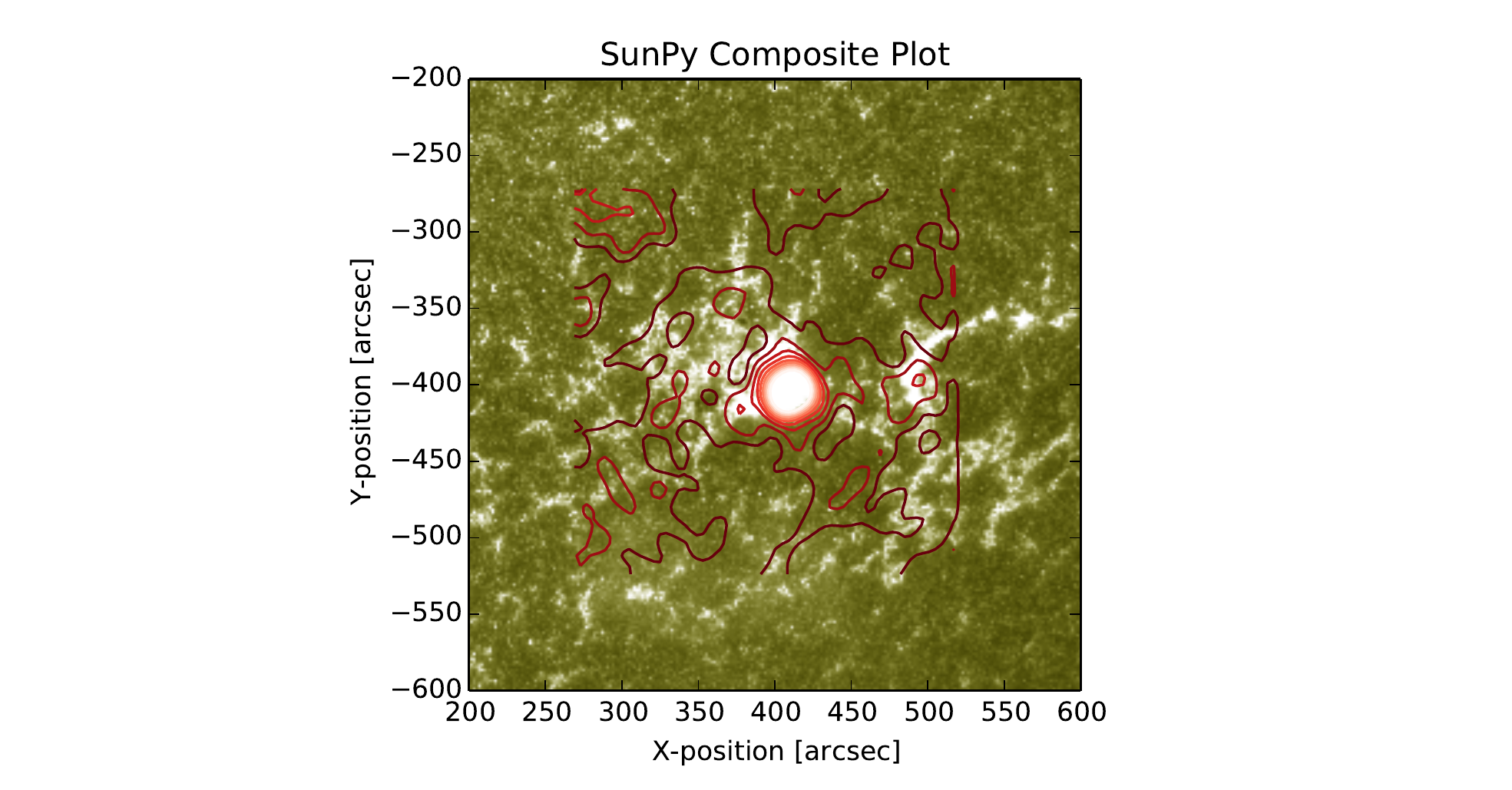}
\end{center}
\caption{Example showing the functionality of \texttt{CompositeMap}, with RHESSI X-ray image data composited
on top of an \textit{SDO}/AIA 1600 $\AA$ image. The \texttt{CompositeMap} is plotted using the integration with the \texttt{matplotlib.pyplot} interface.}
\label{code:compmap_1}
\end{listing}

\begin{listing}[H]
\pythoncode{pycode_map3.txt}
\begin{center}
\includegraphics[width=0.8\columnwidth]{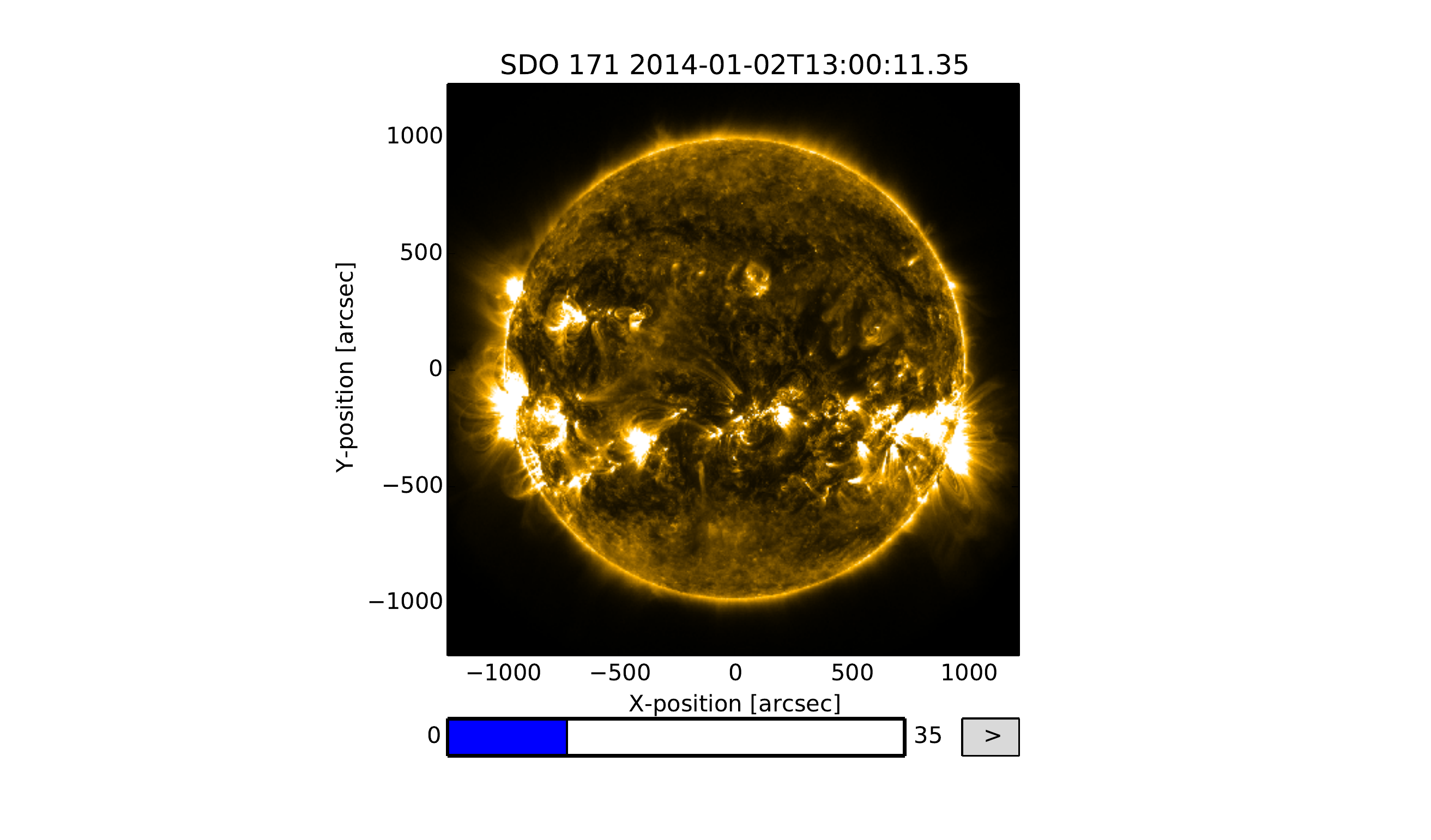}
\end{center}
\caption{Example showing the creation of a \texttt{MapCube} from a list of AIA image files. The 
resultant plot makes use of \texttt{matplotlib}'s interactive widgets to allow scrolling 
through the \texttt{MapCube}.}
\label{code:mapcube_1}
\end{listing}

\subsection{Lightcurve}\label{ssec:lightcurve}

Time series data and their analyses are a fundamental part of solar
physics for which many data sources are available.
SunPy provides a \texttt{LightCurve} class
with a convenient and consistent interface for handling solar time-series
data.  The main engine behind the \texttt{LightCurve} class is
the {\texttt{pandas}} data analysis library.  
\texttt{LightCurve}'s \texttt{data} attribute is a \texttt{pandas.DataFrame} 
object. The \texttt{pandas} library contains a large amount
of functionality for manipulating and analysing time-series data,
making it an ideal basis for \texttt{LightCurve}.  \texttt{LightCurve}
assumes that the input data are time-ordered list(s) of numbers, and each
list becomes a column in the \texttt{pandas} DataFrame object.

Currently, the \texttt{LightCurve} class is compatible with the
following data sources: the Geostationary Operational Environmental
Satellite (\textit{GOES}) X-ray Sensor (XRS), the \textit{Nobeyama
  Radioheliograph (NoRH)}, \textit{PROBA2} Large Yield Radiometer
(LYRA, \citealt{2013SoPh..286...21D}), \textit{RHESSI},
\textit{SDO} EUV Variability Experiment\footnote{Note that only the level ``OCS'' and average
  CSV files is currently implemented -- see
  \url{http://lasp.colorado.edu/home/eve/data/}} (EVE, \citealt{2012SoPh..275..115W}). 
\texttt{LightCurve}
also supports a number of solar summary indices - such as average
sunspot number - that are provided by the National Oceanic and
Atmospheric Administration (NOAA).  For each of these sources, a
subclass of the \texttt{LightCurve} object is initialised (e.g.,
\texttt{GOESLightCurve}) which inherits from \texttt{LightCurve}, but
allows instrument-specific functionality to be included.  Future
developments will introduce support for additional instruments and
data products, as well as implementing an interface similar to that of
\texttt{Map}.  Since there is no established standard as to how
time-series data should be stored and distributed, each SunPy
\texttt{LightCurve} object subclass provides the ability to download
its corresponding specific data format in its constructor and parse
that file type. A more general download interface is currently in development.

A \texttt{LightCurve} object may be created using a number of different methods. 
For example, a \texttt{LightCurve} may be created for a specific instrument given
an input time range. In Listing~\ref{code:goes_lc}, 
the \texttt{LightCurve} constructor searches a remote source for the GOES X-ray 
data specified by the time interval, downloads the required files, and 
subsequently creates and plots the object. Alternatively, if the data file 
already exists on the local system, the \texttt{LightCurve} object may be 
initialised using that file as input.

\begin{listing}[H]
\pythoncode{pycode_lightcurve.txt}
\begin{center}
\includegraphics[width=10cm]{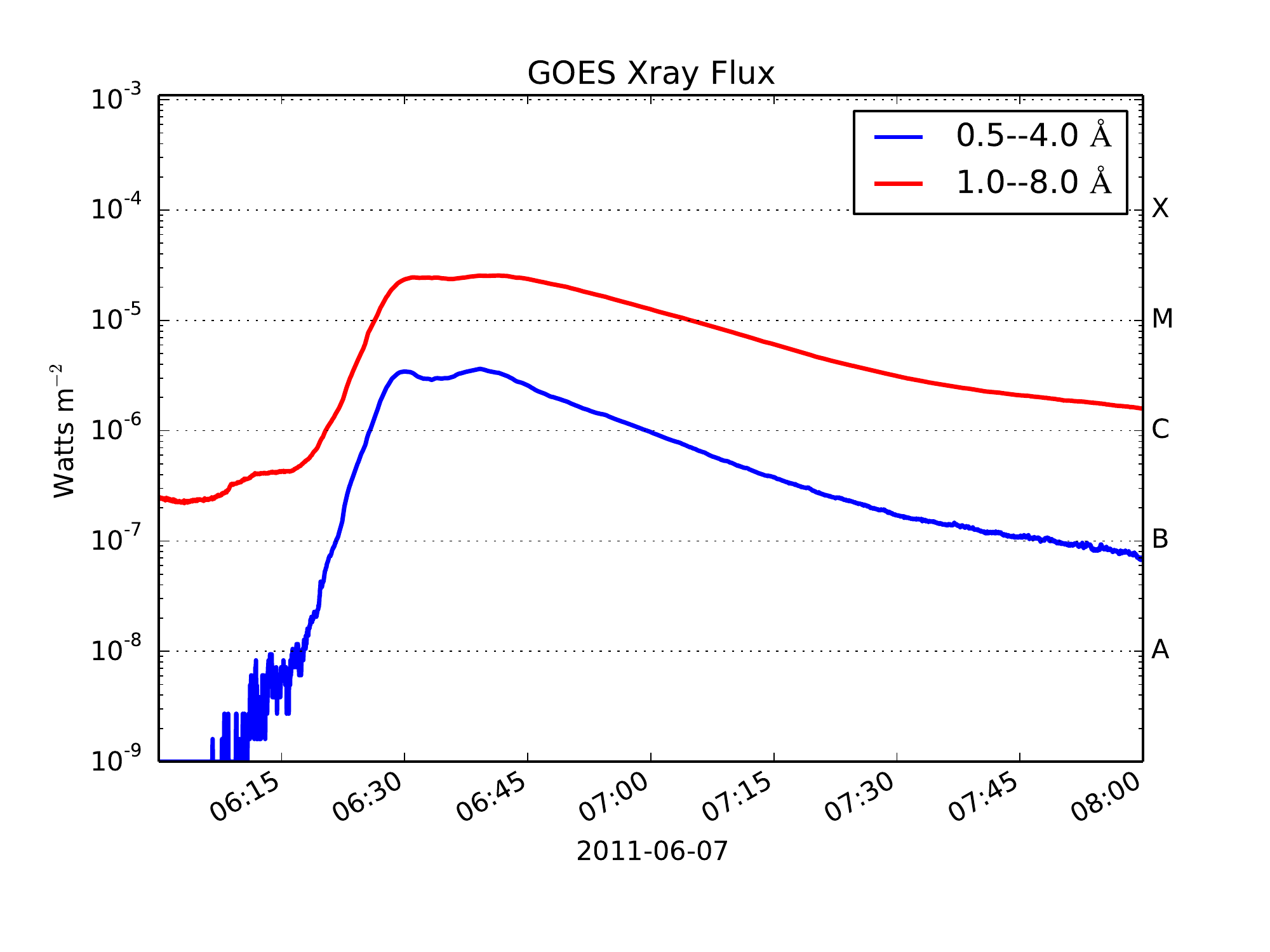}
\end{center}
\caption{Example retrieval of a GOES lightcurve
using a time range and the output of the 
\texttt{peek()} method. The maximum flux value in the \textit{GOES} 1.0--8.0$\AA$\ channel 
is then retrieved along with the location in time of the maximum.}
\label{code:goes_lc}
\end{listing}

\subsection{Spectra}\label{sec:spectra}
SunPy aims to provide broad support for solar spectroscopy
instruments.  The variety and complexity of these instruments and
their resultant datasets makes this a challenging goal.  The \texttt{spectra} module implements a
\texttt{Spectrum} class for 1D data (intensity as a function of frequency) and a
\texttt{Spectrogram} class for 2D data (intensity as a function of time and
frequency).  Each of these classes uses a \texttt{numpy.ndarray} object
as its \texttt{data} attribute.  

As with other SunPy data types, the \texttt{Spectrogram} class has been
built so that each instrument initialises using a subclass containing the instrument-specific 
functionalities. The common functionality provided by the base \texttt{Spectrogram} class includes
joining different time ranges and frequencies, performing frequency-dependent background subtraction,
and convenient visualization and sampling of the data.
Currently, the \texttt{Spectrogram} class supports radio spectrograms from the e-Callisto (
\url{http://www.e-callisto.org/})
solar radio spectrometer network \citep{2009EM&P..104..277B} and \textit{STEREO}/SWAVES spectrograms \citep{2008SSRv..136..487B}.

Listing \ref{code:spectra} shows how the \texttt{CallistoSpectrogram}
object retrieves spectrogram data in the time range specified.  
When the data is requested using the
\texttt{from\_range()} function, the object merges all the downloaded
files into a single spectrogram, across time and frequency.
In the example shown, data is provided in two frequency ranges:
20--90\,MHz and 55--355\,MHz.  Since the data are not evenly spaced in
the frequency range, the \texttt{Spectrogram} object linearises the
frequency axis to assist analysis.  The example also demonstrates
the implemented background subtraction method, which calculates
a constant background over time for each frequency channel.

\begin{listing}[H]
\pythoncode{pycode_spectra.txt}
\begin{center}
\includegraphics[width=0.8\columnwidth]{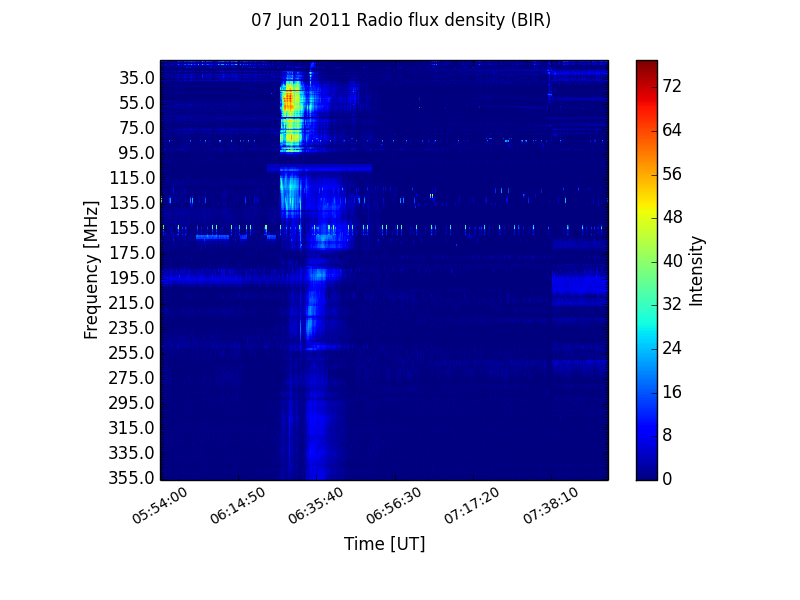}
\end{center}
\caption{Example of how \texttt{CallistoSpectrogram} retrieves the
  data for the requested time range and observatory, merges it, and
  removes the background signal.  The data requested -- `BIR' -- is
  the code name of the Rosse Observatory \url{http://www.rosseobservatory.ie}
  at Birr Castle in Ireland.}
\label{code:spectra}
\end{listing}

\section{Solar Data Search and Retrieval}\label{sec:retrieval}

Several well-developed resources currently exist which provide remote access to 
and data retrieval form a large number of solar and heliospheric data sources 
and event databases. SunPy provides support for these resources via the 
\texttt{net} subpackage. In the following subsections, we describe each of 
these resources and how to use them.

\subsection{VSO}\label{ssec:vso}

The Virtual Solar Observatory (VSO, \url{http://virtualsolar.org}) provides a 
single, standard query interface to solar data from many different archives 
around the world \citep{hill2009}.
Data products can be requested for specific instruments or missions and
can also be requested based on physical parameters of the data product such
as the wavelength range.
In addition to the VSO's primary web-based interface, a SOAP (Simple Object 
Access Protocol) service is also available.
SunPy's \texttt{vso} module provides access to the VSO via this SOAP service using the
\texttt{suds} package.

Listing~\ref{code:vso_query_simple} shows an example of how to query and download data
from the VSO using the \texttt{vso} module.
Queries are constructed using one or more attribute objects. Each
attribute object is a constraint on a parameter of the data set, such as the
time of the observation, instrument, or wavelength.
Listing~\ref{code:vso_query_simple} also shows how to download the data using
the constructed query. The path to which the data files will be downloaded is defined using custom tokens
which reference the file metadata (e.g., instrument, detector, filename). This provides
users the ability to organize their data into subdirectories on download.

Listing~\ref{code:vso_query_advanced} shows an example of how to make an advanced
query by combining attribute objects.
Two attribute objects can be combined with a logical \texttt{or} operation
using the \texttt{|} (pipe) operator.
All attribute objects provided to the query as arguments are combined with a 
logical \texttt{and} operation.

\begin{listing}[H]
\pythoncode{pycode_vso1.txt}
\caption{Example of querying a single instrument over a time range and downloading the data}
\label{code:vso_query_simple}
\end{listing}

\begin{listing}[H]
\pythoncode{pycode_vso2.txt}
\caption{Example of an advanced VSO query using attribute objects,
combining both data from a detector and any data that falls within two wavelength ranges,
continuing from Listing~\ref{code:vso_query_simple}.}
\label{code:vso_query_advanced}
\end{listing}

\subsection{HEK}\label{ssec:hek}

The Sun is an active star and exhibits a wide range of transient phenomena 
(e.g., flares, radio bursts, coronal mass ejections) at many different time-scales, 
length-scales, and 
wavelengths. Observations and metadata concerning these phenomena are collected 
in the Heliophysics Event Knowledgebase (HEK, \citealp{hurlburt2012}).  Entries are generated both by
automated algorithms and human observers.  Some of the information in the HEK 
reproduces feature and event data from elsewhere (for example, the \textit{GOES} flare catalogue),
and some is generated by the Solar Dynamics Observatory Feature Finding Team 
\citep{martens2012}.  A key feature of the HEK is that it
provides an homogeneous and well-described interface to a large amount of 
feature and event information. SunPy 
accesses this information through the \texttt{hek} module.  The \texttt{hek} module makes use of the 
HEK public API\footnote{For more information see \url{http://vso.stanford.edu/hekwiki/ApplicationProgrammingInterface}}.

Simple HEK queries consist of start time, an end time, and an event type 
(see Listing~\ref{code:hek:simple}). Event types are specified as upper case, 
two letter strings, and these strings are 
identical to the two letter abbreviations defined by HEK 
(see \url{http://www.lmsal.com/hek/VOEvent_Spec.html}). Users can see a
complete list and description of these abbreviations by looking at the documentation
for \texttt{hek.attrs.EventType}.

\begin{listing}[H]
\pythoncode{pycode_hek.txt}
\caption{Example usage of the \texttt{hek} module showing a simple HEK search for solar flares
on 2011 August 9.}
\label{code:hek:simple}
\end{listing}

Short-cuts are also provided for some often-used event types. For example, 
the flare attribute can be declared as either 
\texttt{hek.attrs.EventType("FL")} or as \texttt{hek.attrs.FL}. 
HEK attributes differ from VSO attributes (Section \ref{ssec:vso}) in that many 
of them are wrappers that conveniently expose comparisons by overloading Python 
operators. This allows filtering of the HEK entries by the properties of the 
event. As was mentioned above, the HEK stores feature and event metadata obtained 
in different ways, known generally as feature recognition methods (FRMs). 
The example in Listing~\ref{code:hek:frm} repeats the previous 
HEK query (see Listing \ref{code:hek:simple}), with an additional filter enabled 
to return only those events that have the FRM `SSW Latest Events'.  
Multiple comparisons can be made by including more comma-separated
conditions on the attributes in the call to the HEK query method.

\begin{listing}[H]
\pythoncode{pycode_hek2.txt}
\caption{An HEK query that returns only those flares that were
  detected by the `SSW Latest Events' feature recognition method.}
\label{code:hek:frm}
\end{listing}

HEK comparisons can be combined using Python's logical operators (e.g., \texttt{and}
and \texttt{or}). The ability to use comparison and logical operators on HEK attributes allows 
the construction of queries of arbitrary complexity.
For the query in Listing \ref{code:hek:or} returns
returns flares with helio-projective $x$-coordinates west of 50 arcseconds or 
those that have a peak flux above 1000.0 (in units defined by the FRM).

\begin{listing}[H]
\pythoncode{pycode_hek3.txt}
\caption{HEK query using the logical \texttt{or} operator.}
\label{code:hek:or}
\end{listing}
All FRMs report their required feature attributes (as defined by the
HEK), but the optional attributes are FRM dependent\footnote{See
  \url{http://www.lmsal.com/hek/VOEvent_Spec.html} for a list of
  features and their attributes.}.  If a FRM does not have one of the
optional attributes, \texttt{None} is returned by the \texttt{hek}
module.
 
After users have found events of interest the next step is to 
download observational data. The \texttt{H2VClient} module makes this
easier by providing a translation layer between HEK query results
and VSO data queries. This capability is demonstrated in Listing~\ref{code:hek2vso}.
\begin{listing}[H]
\pythoncode{pycode_hek4.txt}
\caption{Code snippet continuing from Listing~\ref{code:hek:or} showing the 
query and download of data from the first HEK result from the VSO.}
\label{code:hek2vso}
\end{listing}

\subsection{HELIO}\label{ssec:helio}

The HELiophysics Integrated Observatory (HELIO)\footnote{For more information 
see \url{http://helio-vo.eu}} has 
compiled a list of web services which allows scientists to query and 
discover data throughout the heliosphere, from solar and magnetospheric data to planetary and 
inter-planetary data \citep{perez-suarez2012}.
HELIO is built with a Service-Oriented Architecture, 
i.e., its capabilities are divided into a number of tasks that are 
implemented as separate services. 
HELIO is made up of nine different public services, 
which allows scientists to search different catalogues of registered events, 
solar features, data from instruments in the heliosphere, and other information 
such as planetary or spacecraft position in time. 
Additionally, HELIO provides a service that uses a 
propagation model to link the data in different points of the solar system by 
its original nature (e.g., Earth auroras are a signature of magnetic 
field disturbances produced a few days before on the Sun).
In addition to the primary, web-based interface to 
HELIO, its services are available via an API.

SunPy's \texttt{hec} module provides an interface to the
HELIO Event Catalogue (HEC) service. 
This module was developed as
part of a Google Summer of Code (GSOC) project in 2013.
The HEC service currently provides access to 84 catalogues from different
sources.
As with all of the HELIO services, the HEC service provides results in VOTable 
data format (defined by IVOA, see \citealt{ochsenbein2011}).
The \texttt{hec} module parses this output using the \texttt{astropy.io.votable} package.
This format has the advantage of containing metadata with information like
data provenance and the performed query.

For example, Listing~\ref{code:helio} shows how to obtain information
from different catalogues of coronal mass ejections (CMEs).

\begin{listing}[h]
\pythoncode{pycode_helio.txt}
\caption{Example of querying the HEC service to multiple CME
catalogues, in this case the ones detected automatically 
by the by the Computer Aided CME Tracking feature recognition algorithm
\citep[CACTus - \url{http://sidc.oma.be/cactus/};][]{robbrecht_automated_2009}.}
\label{code:helio}
\end{listing}
\subsection{Helioviewer}\label{ssec:hv}

SunPy provides the ability to download images hosted by the
Helioviewer Project (\url{http://wiki.helioviewer.org}).  
The aim of the Helioviewer Project is to enable
the exploration of solar and heliospheric data from multiple data
sources (such as instrumentation and feature/event catalogues) via
easy-to-use visual interfaces. The Helioviewer Project have developed two client 
applications that allow users to browse images and create movies of the Sun taken 
by a variety of instruments: \url{http://www.helioviewer.org}, a 
Google Maps-like web application, and \url{http://www.jhelioviewer.org}, 
a movie streaming desktop application. The Helioviewer
project maintains archives of all its image data in JPEG2000 format (\citealt{muller2009}). The
JPEG2000 files are typically highly compressed compared to the source
FITS files from which they are generated, but are still high-fidelity, and thus can be used to quickly
visualise large amounts of data from multiple sources.  SunPy is
also used in Helioviewer production servers to manage the download and
ingestion of JPEG2000 files from remote servers.

The Helioviewer Project categorises image data based on the physical
construction of the source instrument, using a simple hierarchy:
observatory $\rightarrow$ instrument $\rightarrow$ detector
$\rightarrow$ measurement, where ``$\rightarrow$'' means ``provides a''.  
%schriste - say what?!
Each Helioviewer Project JPEG2000 file contains
metadata which are based on the original FITS header
information, and carry sufficient information to permit overlay with
other Helioviewer JPEG2000 files. Images can be accessed either as
PNGs (Section \ref{sssec:hv:png}) or as JPEG2000 files (Section
\ref{sssec:hv:jp}).

\subsubsection{Download a PNG file}\label{sssec:hv:png}

The Helioviewer API allows composition and overlay of images from
multiple sources, based on the positioning metadata in the source FITS
file.  SunPy accesses this overlay/composition capability through the
\texttt{download\_png()} method of the Helioviewer client.  Listing~\ref{code:hv:overlaid}
gives an example of the composition of three
separate image layers into a single image.

\begin{listing}[H]
\pythoncode{pycode_helioviewer.txt}
\begin{center}
\includegraphics[width=0.6\columnwidth]{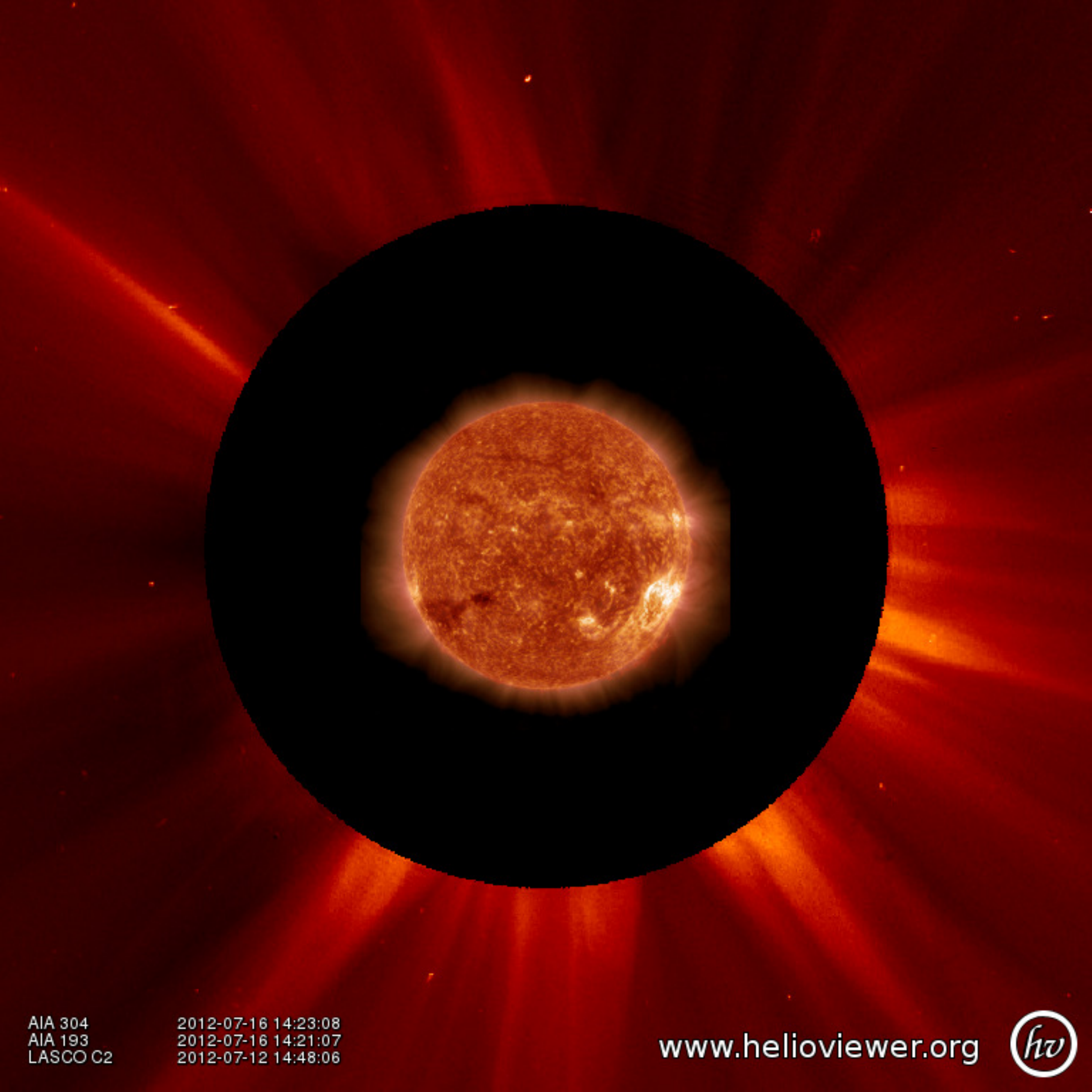}
\end{center}
\caption{Acquisition of a PNG image composed from data from three
  separate sources.}
\label{code:hv:overlaid}
\end{listing}

The first argument is the requested time of the image, and Helioviewer
selects images closest to the requested time.  In this case, the
requested time is in the future and so Helioviewer will find the most
recent available images from each source.  The second argument refers
to the image resolution in arcseconds per pixel (larger values mean
lower resolution).  The third argument is a comma-delimited string of
the three requested image layers, the details of which are enclosed
in parentheses. The image layers are described using the observatory
$\rightarrow$ instrument $\rightarrow$ detector $\rightarrow$
measurement combination described above, along with two following
numbers that denote the visibility and the opacity of the image layer,
respectively (1/0 is visible/invisible, and opacity is in the range
$0\rightarrow100$, with $100$ meaning fully opaque).  The quantities
\texttt{x0} and \texttt{y0} are the $x$ and $y$ centre points about
which to centre the image (measured in helio-projective cartesian
coordinates), and the \texttt{width} and \texttt{height} are the pixel
values for the image dimensions.

This functionality makes it simple for SunPy users to generate complex
images from multiple, correctly overlaid, image data sources.

\subsubsection{Download a JPEG2000 file}\label{sssec:hv:jp}

As noted above, Helioviewer JPEG2000 files contain metadata that allow
positioning of the image data.  There is sufficient metadata in each
file to permit the creation of a SunPy \texttt{Map} object (see Section
\ref{ssec:map}) from a Helioviewer JPEG2000 file.  This allows image
data to be manipulated in the same way as any other map object.

Reading JPEG2000 files into a SunPy session requires installing two
other pieces of software. The first, OpenJPEG
(\url{http://www.openjpeg.org}), is an open-source
library for reading and writing JPEG2000 files.  The other package 
required is Glymur (
\url{https://github.com/quintusdias/glymur}), an
interface between Python and the OpenJPEG libraries (note that these
packages are {\it not} required to use the functionality described in
Section \ref{sssec:hv:png}).

Listing~\ref{code:downloadjp2} demonstrates the querying, downloading,
reading and conversion of a Helioviewer JPEG2000 file into a SunPy map
object.  This functionality allows users to visualise and manipulate
Helioviewer-supplied image data in an identical fashion to a SunPy \texttt{Map}
object generated from FITS data (see Section~\ref{ssec:map}).

\begin{listing}[H]
\pythoncode{pycode_helioviewer2.txt}
\begin{center}
\includegraphics[width=0.8\columnwidth]{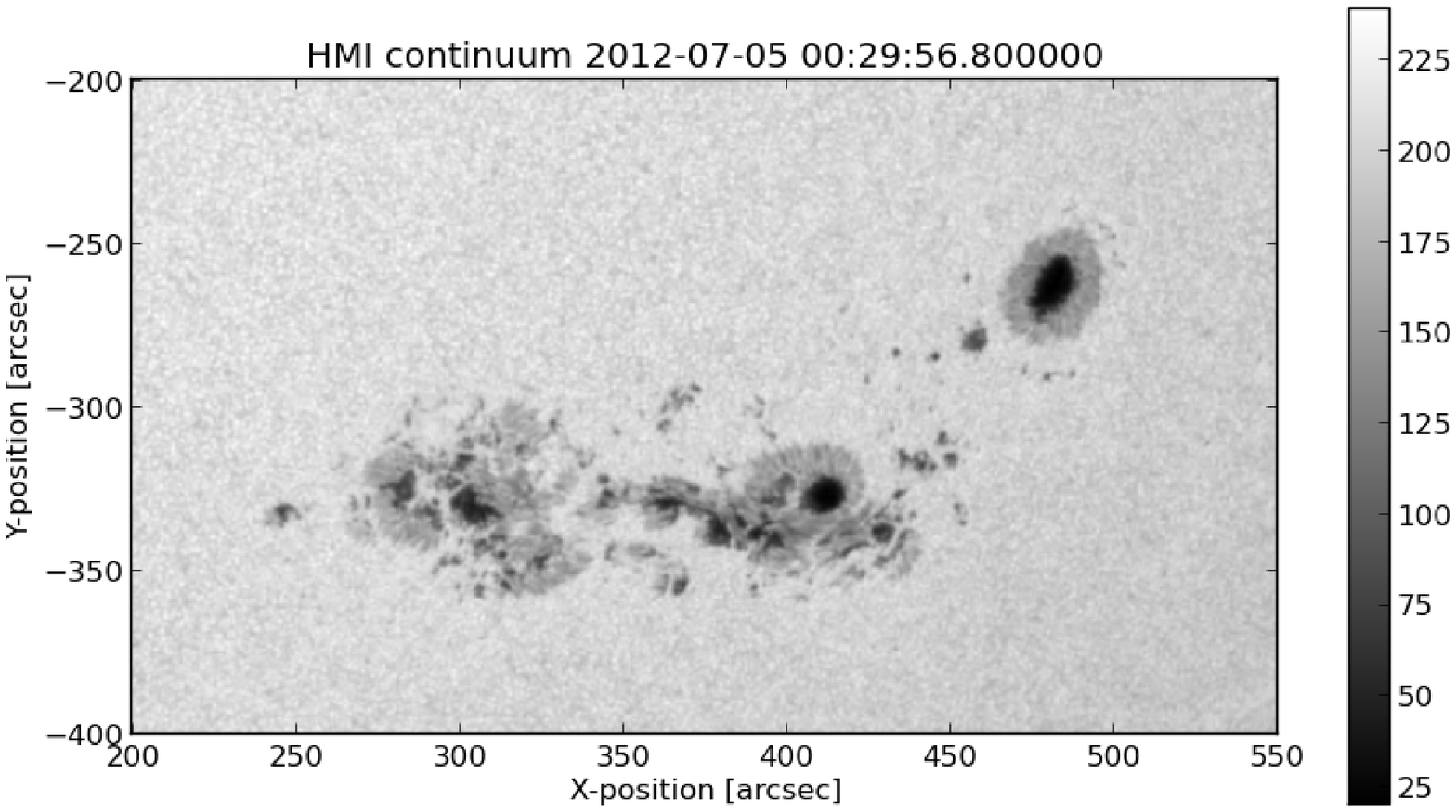}
\end{center}
\caption{Acquisition and display of a Helioviewer JPEG2000 file as a
  SunPy \texttt{Map} object. Images values are byte-scaled in the range 0--255.}
\label{code:downloadjp2}
\end{listing}

\subsection{The File Database}\label{ssec:db}

Easy access to large quantities of solar data frequently leads to data files accumulating
in local storage such as laptops and desktop computers. Keeping data organised and available
is typically a cumbersome task for the average user. The file database is a subpackage of 
SunPy that addresses this problem by providing a unified database to store and 
manage information about local data files.

The \texttt{database} subpackage can make use of any database software supported by
\texttt{SQLAlchemy} (\url{http://www.sqlalchemy.org}). This library was chosen
since it supports many SQL dialects. 
If SQLite is selected, the database is stored as a single file, which is
created automatically. A server-based database, on the other hand, could be used
by collaborators who work together on the
same data from different computers: a central database server stores all data and the clients connect to
it to read or write data.

The database can store and manage all data that can be read via SunPy's 
\texttt{io} subpackage, and direct integration with the \texttt{vso} 
module is supported.
It is also possible to manually add file or directory entries. The package also provides
a unified data search via the \texttt{fetch()} method, which includes both local files
and files on the \textsc{VSO}. This reduces the likelihood of downloading the same file 
multiple times. When a file is added to the database, the file is scanned for metadata,
and a file hash is produced. 
The current date is associated with the entry along with metadata summaries such 
as instrument, date of observation, field of view, etc. 
The database also provides the ability to associate custom metadata to 
each database entry such as keywords, comments, and favourite tags, as well as 
%dps: favouring is not what we want to say here...
%JI: ``assigning favourites'' - is this what is meant?
%dps: Yes, how does it sounds now?
querying the full metadata (e.g., FITS header) of each entry.

The \texttt{Database} class connects to a database and allows the user to 
perform operations on it. Listing~\ref{code:database} shows how to connect
to an in-memory database and download data from the \textsc{VSO}. These entries are
automatically added to the database. The function \texttt{len()} is used to get the number of
records. The function \texttt{display\_entries()} displays an iterable of 
database entries in a formatted \textsc{ASCII} table. The headlines 
correspond to the attributes of the respective database entries.

A useful feature of the database package is the support of \texttt{undo}
and \texttt{redo} operations. This is particularly convenient in
interactive sessions to easily revert accidental operations. 
This feature will also be desirable for a planned GUI frontend for this package.

\begin{listing}[H]
\pythoncode{pycode_database.txt}
\caption{Example usage of the \texttt{database} subpackage.}
\label{code:database}
\end{listing}

\section{Additional Functionality}\label{sec:util}
SunPy is meant to provide a consistent environment for solar data analysis. In 
order to achieve this goal SunPy provides a number of additional functions and packages which 
are used by the other SunPy modules and are made available to the user. This section 
briefly describes some of these functions.
	
\subsection{World Coordinate System (WCS) Coordinates}\label{ssec:util:wcs}
Coordinate transformations are frequently a necessary task within the solar 
data analysis workflow. An often used transformation is from 
observer coordinates (e.g., sky coordinates) to a coordinate system that is 
mapped onto the solar surface (e.g., latitude and longitude). This 
transformation is necessary to compare the true physical distance between 
different solar features. This type of transformation is not unique
to solar observations, but is not often considered by astronomical packages
such as the Astropy 
\texttt{coordinates} package. The \texttt{wcs} package in SunPy implements the World Coordinate 
System (WCS) for solar coordinates as described by \cite{thompson2006}. The 
transformations currently implemented are some of the most commonly used in solar data analysis, namely converting from Helioprojective-Cartesian 
(HPC) to Heliographic (HG) coordinates. HPC describes the positions on 
the Sun as angles measured from the center of the solar disk (usually in 
arcseconds) using Cartesian coordinates (X, Y). This is the coordinate system 
most often defined in solar imaging data (see for example, images from 
\textit{SDO}/AIA, \textit{SOHO}/EIT, and \textit{TRACE}). 
HG coordinates express positions on the Sun using longitude and latitude on 
the solar sphere. There are two standards for this coordinate system:
Stonyhurst-Heliographic, where the origin is at the intersection of the solar 
equator and the central meridian as seen from Earth, and 
Carrington-Heliographic, which is fixed to the Sun and does not depend on Earth. The 
implementation of these transformations pass through a common coordinate system 
called Heliocentric-Cartesian (HCC), where positions are expressed in true 
(de-projected) physical distances instead of angles on the celestial sphere.
These transformations require some knowledge of the location of the observer, 
which is usually provided by the image header. In the cases where it is 
not provided, the observer is assumed to be at Earth. Listing \ref{code:wcs_code} shows 
some examples of coordinate transforms carried out in SunPy using the 
\texttt{wcs} utilities. This will form the foundation for transformations functions
to be used on \texttt{Map} objects.

\begin{listing}[H]
\pythoncode{pycode_util1.txt}
\caption{Using the \texttt{wcs} subpackage.}
\label{code:wcs_code}
\end{listing}

\subsection{Solar Constants and units}\label{ssec:util:sun}
Physical quantities (i.e. a number associated with a unit) are an important part
of scientific data analysis. SunPy makes use of the \texttt{Quantity} object provided by 
Astropy \texttt{units} sub-package. This object maintains the relationship between 
a number and its unit and makes it easy to convert between units. 
As these objects inherit from 
NumPy's \texttt{ndarray}, they work well with standard representations of numbers. Using proper
quantities inside of the code base also makes it easier to catch errors in calculations.
SunPy is currently working on integrating quantities throughout the code base.
In order to encourage the use of units and to enable consistency SunPy provides
the \texttt{sun} subpackage which includes solar-specific data such as ephemerides and
solar constants. The main namespace contains a number of functions that provide solar
ephemerides such as the Sun-to-Earth distance, solar-cycle number, mean 
anomaly, etc.
All of these functions take a time as their input, which can be provided in a format
compatible with \texttt{sunpy.time.parse\_time()}. 

The \texttt{sunpy.sun.constants} module provides a number of solar-related 
constants in order to enable the calculation of derived solar 
values within SunPy, but also to the user. All solar 
constants are provided as \texttt{Constant} objects as defined in the Astropy \texttt{units} package. Each 
\texttt{Constant} object defines a \texttt{Quantity}, along with 
the constant's provenance (i.e., reference) and its uncertainty. The use of this package
is shown in Listing~\ref{code:constants_code}.
For convenience, a number of shortcuts to frequently used constants are provided 
directly when importing the module. A larger list of constants can be 
accessed through an interface modeled on that provided by the SciPy \texttt{constants}
package and is available as a dictionary called \texttt{physical\_constants}. 
To view them all quickly, a \texttt{print\_all()} function is available.

\begin{listing}[H]
\pythoncode{pycode_util2.txt}
\caption{Using the \texttt{sun.constants} module.}
\label{code:constants_code}
\end{listing}
	
\subsection{Instruments}\label{ssec:util:inst}
In addition to providing support for instrument-specific solar data via the main data 
classes \texttt{Map}, \texttt{LightCurve}, and \texttt{Spectrum}, 
some instrument-specific functions may be found within the \texttt{instr} subpackage. 
These functions are generally those that are unique to one particular solar instrument, 
rather than of general use, such as a function to construct a \textit{GOES} flare event list 
or a function to query the \textit{LYRA} timeline annotation file. Currently, some support is included
for the \textit{GOES}, \textit{LYRA}, \textit{RHESSI} and \textit{IRIS} instruments, while future developments 
will include support for additional missions. Ultimately, it is anticipated that solar
missions requiring a large suite of software tools will each be supported via a separately 
maintained package that is affiliated with SunPy.

\section{Development and Community}\label{sec:dev}
SunPy is a community-developed library, designed and developed for and by 
the solar physics community. Not only is all the source code publicly available 
online under the permissive 2-clause BSD licence, the whole 
development process is also online and open for anyone to contribute to.
SunPy's development makes use of the online service 
GitHub (\url{http://github.com}) and Git\footnote{For more information see \url{http://git-scm.com/}}
as its distributed version control software. 

The continued success of an open-source project depends on many factors;
three of the most important are (1) utility and quality of the code, (2) documentation, and (3) an
active community \citep{bangerth2013}. Several tools, some specific to Python, are used by
SunPy to make achieving these goals more accessible. To maintain high-quality code, a 
transparent and collaborative development workflow made possible by GitHub is used.
The following conditions typically must be met before code is accepted.
\begin{enumerate}
	\item  The code must follow the
	PEP 8 Python style 
	guidelines (\url{http://www.python.org/dev/peps/pep-0008/}) to maintain consistency in the SunPy code.
	
	\item All new features require documentation in the form of doc strings as well as user
	guides. 
	
	\item The code must contain unit tests to verify that the code is behaving 
	as expected.

    \item Community consensus is reached that the new code is valuable and appropriately implemented.
\end{enumerate}
This kind of development model is widely used within the scientific Python 
community as well as by a wide variety of other projects, both open and closed 
source. 

Additionally, SunPy makes use of `continuous integration' provided by Travis CI (\url{http://travis-ci.org}), a process by which the addition of any new code 
automatically triggers a comprehensive review of the code functionality which are maintained as unit tests.
 If any single test
fails, the community is alerted before the code is accepted. The unit-test coverage is monitored by
a service called Coveralls (\url{http://coveralls.io}).

High-quality documentation is
one of the most important factors determining the success of any software project. 
Powerful tools already exist in Python to support documentation, thanks to native
Python's focus on its own documentation. SunPy makes use of the Sphinx (\url{http://sphinx-doc.org})
documentation generator. Sphinx uses reStructuredText as its markup language, which is
an easy-to-read, what-you-see-is-what-you-get plaintext markup syntax. It supports
many output formats most notably HTML, as well as PDF and ePub, and provides a rich,
hierarchically structured view of in-code documentation strings. The SunPy documentation 
is built automatically and is hosted by Read-the-Docs (\url{http://readthedocs.org})
at \url{http://docs.sunpy.org}. 

Communication is the key to maintaining an active community, and the SunPy community 
uses a number of different tools to facilitate communication. For immediate communications, an active IRC chat
room (\#SunPy) is hosted on freenode.net. For more involved or less immediate needs, such as
developer comments or discussions, an open mailing list is hosted by Google Groups. 
Bug tracking, code reviews, and feature-request discussions take place directly on GitHub.
The SunPy community also reaches out to the wider solar physics
community through presentations, functionality demonstrations, and informal meetups at scientific
meetings. 

In order to enable the long-term development of SunPy, a formal organizational
structure has been defined. The management of SunPy is the responsibility of 
the SunPy board, a group of elected members of the community. The board elects
a lead developer whose is responsible for the day to day development of SunPy.
SunPy also makes use of Python-style Enhancement proposals which can be proposed
by the community and are voted on by the board. These proposals set the overal
direction of SunPy's development.

\section{Future of SunPy}\label{sec:future}

Over the three years of SunPy's development, the code base has grown to over 17,000 lines. 
SunPy is already a 
useful package for the analysis of calibrated solar data, and it
continues to gain significant new capabilities with each successive release.
The primary focus of the 
SunPy library is the analysis and visualisation of `high-level' solar 
data. This means data that has been put through instrument processing 
and 
calibration routines, and contains valid metadata. 
The plan for SunPy is to continue development within this 
scope. The 
primary components of this plan are to provide a set of data types 
that are 
interchangeable with one another: e.g., if you slice a 
\texttt{MapCube} 
along one spatial location, a \texttt{LightCurve} of intensity along the 
time range of 
the \texttt{MapCube} should be returned. To achieve this goal, all the 
data 
types need to share a unified coordinate system architecture so that 
each data 
type is aware of what the physical type of its data is and how 
operations on 
that data should be performed. This will enable useful operations
such as the coordinate and solar-rotation-aware 
overplotting of HELIO (Section \ref{ssec:helio}) and HEK
results (Section \ref{ssec:hek}) onto maps (Section \ref{ssec:map}).
Finally, support for new data providers and services will be integrated into SunPy.
For example, new HELIO services will be supported by SunPy, aiming for
seamless interaction between the other services and tools available (e.g., 
\texttt{hek}, \texttt{map}).  

In concert with the work on the data types, further integration with 
the 
\texttt{astropy} package will enable SunPy to incorporate many new features
with little effort. Collaboration and joint development with the 
Astropy project \citep{theastropycollaboration2013} is ongoing.

\section{Summary}
We have presented the release of SunPy (v0.5), a Python package for solar physics. In
this paper we have described the main functionality which includes the SunPy data types, 
\texttt{Map} (see Section~\ref{ssec:map}), \texttt{Lightcurve} (see Section~\ref{ssec:lightcurve}), and \texttt{Spectrogram} (see Section~\ref{sec:spectra}).
We have described the data and event catalogue retrieval capabilities of SunPy for
the Virtual Solar Observatory (see Section~\ref{ssec:vso}), the Heliophysics Event Knowledgebase (see Section~\ref{ssec:hek}), as well as
the Heliophysics Integrated Observatory (see Section~\ref{ssec:helio}). We described
a new organization tool for data files integrated into SunPy (see Section~\ref{ssec:db}) 
and we discussed the community aspects, development model (see Section~\ref{sec:dev}), and future plans (see Section~\ref{sec:future}) for the project.
We invite members of the community to contribute to the effort by using SunPy for their 
research, reporting bugs, and sharing new functionality with the project.

\section{Acknowledgements}
Many of the larger features in SunPy have been developed with the generous support of 
external organizations. 
Initial development of SunPy's VSO and HEK implementations were funded by ESA's 
Summer of Code In Space (SOCIS 2011, 2012, 2013) program, as well as a prototype GUI and an
N-dimensional data-type implementation. In 2013, with support from Google's 
Summer Of Code (GSOC) program, through the Python Software Foundation, the 
\texttt{helio}, \texttt{hek2vso}, and \texttt{database} subpackages were 
developed. The Spectra and Spectrogram classes were implemented with support 
from the Astrophysics Research Group at Trinity College Dublin, Ireland, in 
2012.

%%% References
\bibliographystyle{stunat}
\bibliography{Sunpy_paper_0.4}{}

\end{document}